\documentclass[prb,twocolumn,showpacs]{revtex4}

\usepackage{bm}
\usepackage{amsmath}
\usepackage{amssymb}
\usepackage{graphicx}
\usepackage{amsfonts}
\usepackage{txfonts}
\usepackage{graphicx}

\begin{document}

\title{Electric field driven quantum phase transition between band insulator
and topological insulator}
\author{Jun Li}
\affiliation{SKLSM, Institute of Semiconductors, Chinese Academy of Sciences, P.O. Box
912, Beijing 100083, China}
\author{Kai Chang}
\affiliation{SKLSM, Institute of Semiconductors, Chinese Academy of Sciences, P.O. Box
912, Beijing 100083, China}
\date{\today }

\begin{abstract}
We demonstrate theoretically that electric field can drive a quantum
phase transition between band insulator to topological insulator in
CdTe/HgCdTe/CdTe quantum wells. The numerical results suggest that
the electric field could be used as a switch to turn on or off the
topological insulator phase, and temperature can affect
significantly the phase diagram for different gate voltage and
compositions. Our theoretical results provide us an efficient way to
manipulate the quantum phase of HgTe quantum wells.
\end{abstract}

\maketitle

Topological insulator (TI) is a very recent discovery and has attracted a
rapid growing interests due to its novel transport peoperty\cite%
{Kane,BHZ,Marcus,Fu}. Topological insulators possess a gap in the bulk but a
gapless edge states at its boundary, therefore display remarkable transport
property due to the presence of the topological edge states, e.g., quantum
spin Hall effect (QSHE). The QSHE is protected by the time-reversal symmetry
and robust against the local perturbation, e.g., impurity scattering.
Searching for new TI becomes a central issue in this rapid growing field.
Recently, the HgTe QWs have been demonstrated to be a 2D TI to exhibit the
QSHE\cite{Marcus} and BiSb alloys have been proven to be a 3D TI with a
conducting surface\cite{Hsieh}. A few other materials, such as InAs/GaSb
QWs, BiSe, BiTe and SbTe alloys\cite{OtherMat}, are also predicted to be TIs
and demonstrated experimentally. Besides finding new TI materials, searching
the ways to drive a band insulator (BI) into a topological insulator is also
important. It has been demonstrated to be possibly realized by tuning the
thickness of HgTe QW. However, tuning the thickness of QW is not a
convenient way to drive the phase transition. Therefore other efficient ways
such as external fields and temperature is highly desirable to drive a BI
into a TI. These ways would be very important for both potential device
applications and basic physics.

In this Letter, we demonstrate theoretically that a BI can be driven
into a TI by tuning external electric field in
CdTe/Hg$_{1-x}$Cd$_{x}$Te/CdTe QWs based on the self-consistent
calculation of the eight-band Kane model and the Poisson equation.
We demonstrate electric field can change the interband coupling
significantly and consequently leads to strong variations of the
band structures, i.e., therefore leads to the quantum phase
transition from a BI to a TI. We also show phase diagrams at plenty
of parameters and consider the temperature effect on the phase
transition. One can see that the critical gate voltage of external
phase transition can be reduced at high temperature, small Cd
composition and thick thickness of well. Our results could be useful
in finding the optimized parameters to realize the phase transition
in experiment.

\begin{figure}[t]
\includegraphics[width=1\columnwidth]{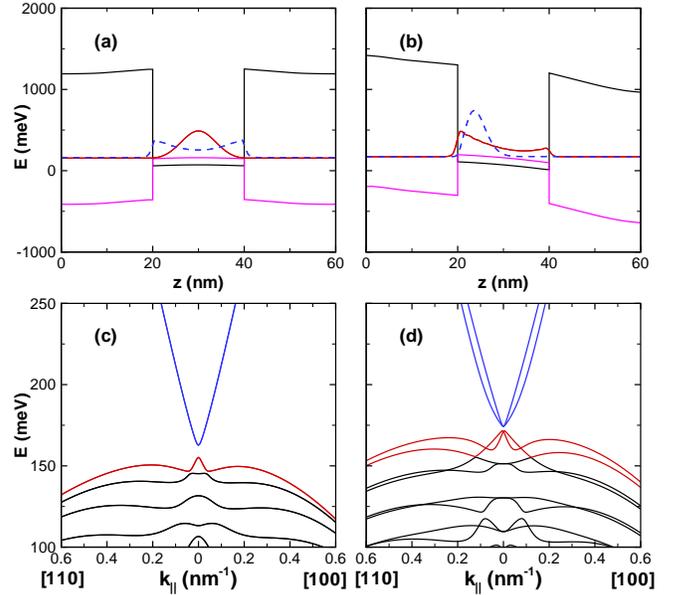}
\caption{(Colour online) Self-consistently calculated band profiles
($T=0$
K) (a) and energy dispersion (c) of a symmetrically doped CdTe (20 nm)/Hg$%
_{0.88}$Cd$_{0.12}$Te (20 nm)/CdTe (20 nm) QW without external gate
voltage. (b) and (d) are the same as (a) and (c) but with gate
voltage $U_{ex}=0.5$ V between the top and bottom gates. The red
solid and blue dashed curves in Fig. 1(a) and 1(b) denote the
probability distributions of the states of the highest valence
subband and lowest conduction subband, respectively. The doping
density is $N_{d}=5\times 10^{11}$ cm$^{-2}$.} \label{fig:fig1}
\end{figure}

We consider a CdTe/Hg$_{1-x}$Cd$_{x}$Te/CdTe QW grown along the [001]
direction [see Fig. \ref{fig:fig1} (a)]. The axis $x$, $y$, and $z$ is
choosen to be along [100], [010], and [001], respectively. Within the
envelope function approximation, the Kane model is a good starting point for
systems with strong interband coupling. When an external voltage is applied
perpendicular to the quantum well plane (at the two sides of the QW),
electrons inside the QW will redistribute due to the effect of the electric
field. The charge redistribution induces an internal electric field, which
affects the charge density distribution, therefore we need to solve the
eight-band Kane model and the Poisson equation self-consistently.

The Kane Hamiltonian for zinc-blende crystals near the $\Gamma $ point can
be written as
\begin{widetext}
\begin{equation}
H_{k}=
\begin{bmatrix}
A & 0 & i\sqrt{3}V^{\dagger} & \sqrt{2}U & iV & 0 & iU & \sqrt{2}V\\
0 & A & 0 & -V^{\dagger} & i\sqrt{2}U & -\sqrt{3}V &
i\sqrt{2}V^{\dagger} &
-U\\
-i\sqrt{3}V & 0 & -(P+Q) & L & M & 0 & \frac{i}{\sqrt{2}}L & -i\sqrt{2}M\\
\sqrt{2}U & -V & L^{\dag} & -(P-Q) & 0 & M & i\sqrt{2}Q & i\sqrt{\frac{3}{2}%
}L\\
-iV^{\dag} & -i\sqrt{2}U & M^{\dag} & 0 & -(P-Q) & -L & -i\sqrt{\frac{3}{2}%
}L^{\dag} & i\sqrt{2}Q\\
0 & -\sqrt{3}V^{\dag} & 0 & M^{\dag} & -L^{\dag} & -(P+Q) &
-i\sqrt{2}M^{\dag}
& -\frac{i}{\sqrt{2}}L^{\dag}\\
-iU & -i\sqrt{2}V & -\frac{i}{\sqrt{2}}L^{\dag} & -i\sqrt{2}Q &
i\sqrt
{\frac{3}{2}}L & i\sqrt{2}M & -P-\Delta & 0\\
\sqrt{2}V^{\dag} & -U & i\sqrt{2}M^{\dag} &
-i\sqrt{\frac{3}{2}}L^{\dag} & -i\sqrt{2}Q & \frac{i}{\sqrt{2}}L & 0
& -P-\Delta
\end{bmatrix}
\label{eqn:Hamil}
\end{equation}
\end{widetext}where
\begin{subequations}
\begin{align}
A& =E_{v}+E_{g}+\boldsymbol{k}A_{c}\boldsymbol{k}, \\
P& =-E_{v}+\frac{\hbar ^{2}}{2m_{0}}\boldsymbol{k}\gamma _{1}\boldsymbol{k},
\\
Q& =\frac{\hbar ^{2}}{2m_{0}}(k_{x}\gamma _{2}k_{x}+k_{y}\gamma
_{2}k_{y}-2k_{z}\gamma _{2}k_{z}), \\
L& =i\frac{\sqrt{3}\hbar ^{2}}{m_{0}}\{k_{-}\gamma _{3}k_{z}\},
\end{align}
\begin{align}
M& =-\frac{\sqrt{3}\hbar ^{2}}{2m_{0}}[k_{x}\gamma _{2}k_{x}-k_{y}\gamma
_{2}k_{y}-2i\{k_{x}\gamma _{3}k_{y}\}], \\
U& =\frac{1}{\sqrt{3}}P_{0}k_{z}, \\
V& =\frac{1}{\sqrt{6}}P_{0}k_{-}.\phantom{\gamma_{2}k_{x}+k_{y}\gamma_{2}
k_{y}-2k_{z}\gamma_{2}k_{z})}  \label{eq:Helements}
\end{align}
where $\boldsymbol{k}=(\boldsymbol{k}_{\parallel },-i\partial /\partial
_{z}) $, $k_{\pm }=k_{x}\pm ik_{y}$, and $\{k_{\alpha }\gamma k_{\beta
}\}=\left( k_{\alpha }\gamma k_{\beta }+k_{\beta }\gamma k_{\alpha }\right)
/2$ $\left( \alpha ,\beta =x,y,z\right) $. Here the in-plane momentum is a
constant of motion and can be replaced by its eigenvalue $\boldsymbol{k}%
_{\parallel }$.The total Hamiltonian becomes $H(\boldsymbol{k}_{\parallel
})=H_{k}(\boldsymbol{k}_{\parallel })-eV_{in}\left( z\right) +eV_{ex}\left(
z\right) $.

The subband dispersions and the corresponding eigenstates are obtained from
the Schr\"{o}dinger equation
\end{subequations}
\begin{equation}
H(\boldsymbol{k}_{\parallel })\left\vert \Psi _{s}(\boldsymbol{k}_{\parallel
})\right\rangle =E_{s}(\boldsymbol{k}_{\parallel })\left\vert \Psi _{s}(%
\boldsymbol{k}_{\parallel })\right\rangle ,  \label{eq:sch}
\end{equation}%
where $s$ is the index of the subband and $\left\vert \Psi _{s}(\boldsymbol{k%
}_{\parallel })\right\rangle =\exp (i\boldsymbol{k}_{\parallel }\boldsymbol{%
\cdot \rho })[\varphi _{1}^{s}(z),\varphi _{2}^{s}(z),...,\varphi
_{8}^{s}(z)]^{T}$ is the envelope function. We solve the Schr\"{o}dinger
equation by expanding $\varphi _{n}^{s}$ by a series of plane waves\cite%
{SpuriousSolution}. In our calculation, $N\approx 30$ is good enough to get
convergent results.

The internal electrostatic potential $V_{in}\left( z\right) $ is determined
by the Poisson equation

\begin{equation}
\frac{d}{dz}\varepsilon (z)\frac{d}{dz}V_{in}(z)=\rho _{h}(z)-\rho
_{e}(z)+N_{d}D\left( z\right) ,  \label{eq:poisson}
\end{equation}%
where $\rho _{e}(z)$ and $\rho _{h}(z)$ are, respectively, the charge
density of electrons and holes and $\varepsilon (z)$ is the static
dielectric constant. $N_{d}$ is the ionized donors density and $D\left(
z\right) $ is the distribution function of the impurities. In this paper, $%
D\left( z\right) $ is assumed to be a exponentially decayed function. For
simplicity, we take $T=0$ K and the axial approximation in the
self-consistent procedure. The Fermi level $E_{F}$ is determined by the
charge neutrality condition $\int_{0}^{L}\left[ \rho _{e}(z)-\rho _{h}(z)%
\right] dz=N_{d}$, where $L=L_{HgCdTe}+2L_{CdTe}$ is the total width
of barriers and QW. The thickness of CdTe barrier $L_{CdTe}$ is
fixed at 20 nm.

While the external electrostatic potential $V_{ex}\left( z\right) $ can be
determined by

\begin{equation}
\frac{d}{dz}\varepsilon (z)\frac{d}{dz}V_{ex}(z)=0,  \label{eq:Vex}
\end{equation}%
and the boundary condition

\begin{equation}
\int_{0}^{L}\varepsilon (z)\frac{d}{dz}V_{ex}(z)=\varepsilon _{0}U_{ex}.
\label{eq:BC}
\end{equation}%
$U_{ex}$ is the external voltage which could be applied on the top and
bottom gate at the two sides of the QW.

In Table \ref{tab:para} we list the band parameters used in our calculation%
\cite{Parameters1,Parameters2}. The Kane parameters of
Hg$_{1-x}$Cd$_{x}$Te can be assumed to be
$x$-independent\cite{Parameters1}, since the band structure
dependence on the Cd composition $x$ caused mainly by the
variation of the band gap $E_{g}$. And the band gap $E_{g}$ of Hg$_{1-x}$Cd$%
_{x}$Te can be obtained in the previous work\cite{Parameters3}.

\begin{table}[ptb]
\centering
\begin{ruledtabular}%
\caption{Band structure parameters of Hg$_{1-x}$Cd$_{x}$Te\cite{Parameters1} and CdTe\cite{Parameters2} used in the Kane Hamiltonian.}%
\begin{tabular}
[c]{lllllllll} & $E_g$ (eV) & $\Delta$ (eV) & $E_p$ (eV) & $F$ &
$\gamma_1$ & $\gamma_2$ & $\gamma_3$ & $\varepsilon$ \\
\hline
Hg$_{1-x}$Cd$_{x}$Te &  & 1.0 & 19.0 & -0.8 & 3.3 & 0.1 & 0.9 & 21 \\
CdTe                 & 1.606 & 0.91 & 18.8 & -0.09 & 1.47 & -0.28 & 0.03 & 10.4 \\

\end{tabular}
\label{tab:para}%
\end{ruledtabular}
\end{table}

In Figs. \ref{fig:fig1} (a) and 1(b) we show the self-consistently
calculated band profiles of CdTe/Hg$_{0.88}$Cd$_{0.12}$Te/CdTe QW
with and without external electric field. The strong interband
coupling makes the quantum states near the gap in
CdTe/Hg$_{1-x}$Cd$_{x}$Te/CdTe QWs very different from those in
conventional semiconductor QWs. This can be seen clearly from the
wavefunction distributions of carriers. For the normal band
structure, the probability of the highest valence subband state
localizes at the center of the QW, while the probability of lowest
conduction subband state is more localized in the vicinity of the
sides of the QW (see the red solid and blue dashed curves in Fig.
\ref{fig:fig1} (a)). This is because the main component of the
lowest conduction subband state for the normal band structure is
electron and it couples strongly with the light-hole component even
at $\mathbf{k}_{\mathbf{\Vert }}=0$, while the main component of the
highest valence subband state is heavy-hole and it decouples with
the other components at $\mathbf{k}_{\mathbf{\Vert }}=0$. When an
external gate voltage is applied, the band structure changes from
the normal band structure to the inverted band structure, and the
probability distributions of the lowest conduction subband and
highest valence subband states exchange
(see the red and blue curves in Fig. \ref{fig:fig1} (b)). In Figs. \ref%
{fig:fig1}(c) and (d) we plot the corresponding band structures of CdTe/Hg$%
_{1-x}$Cd$_{x}$Te/CdT QWs with and without electric field. We demonstrate
clearly that electric field can change the interband coupling significantly
and consequently leads to strong variations of the band structures, i.e.,
the variation from the normal band structure to the inverted band structure.
Due to the enhanced interband coupling, the lowest conduction subbands and
the highest valence subbands exhibit a strong anticrossing behavior and open
a mini hybridized gap at a finite $\boldsymbol{k}_{\parallel }$ in the case
of the inverted band structure. The edge states will appear if a lateral
confinement is applied to this QW with the inverted band structure. Our
numerical results demonstrate that a normal insulator indeed can be driven
into a topological insulator electrically.

\begin{figure}[t]
\includegraphics[width=1\columnwidth]{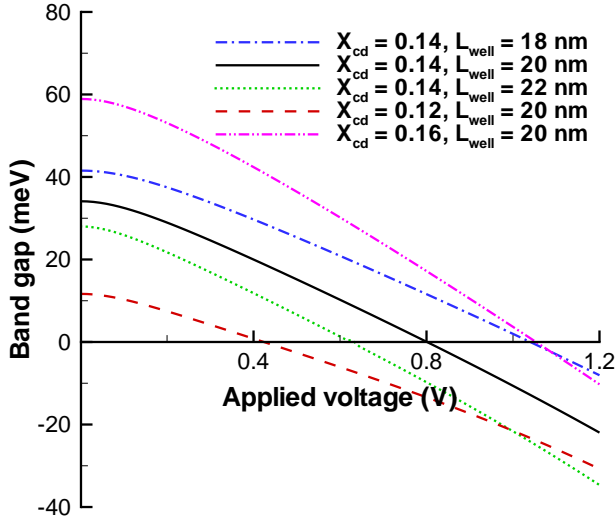}
\caption{(Colour online) Bandgap of CdTe/Hg$_{1-x}$Cd$_{x}$Te/CdTe
QW ($T =
0 $ K) as a function of external gate voltage for different thickness of Hg$%
_{1-x}$Cd$_{x}$Te well and the composition of Cd in
Hg$_{1-x}$Cd$_{x}$Te alloy.} \label{fig:fig2}
\end{figure}

Next, we turn to discuss how electric fields affect the bandgap of QWs. Fig. %
\ref{fig:fig2} shows that the bandgap of
CdTe/Hg$_{1-x}$Cd$_{x}$Te/CdTe QW as a function of the external gate
voltage for QWs with different
thicknesses of well and Cd compositions. In thin CdTe/Hg$_{1-x}$Cd$_{x}$%
Te/CdTe QWs without external electric fields, the quantum confinement effect
can push the lowest conduction subbands to higher energy and therefore the
QW exhibits the normal band structure. From this figure, one can see clearly
that for a certain Cd composition and thickness of well, the bandgap of QW
can be reduced into the negative value by tuning external voltages.
Increasing the Cd composition or decreasing the thickness of the Hg$_{1-x}$Cd%
$_{x}$Te well, the band gap of unbiased CdTe/Hg$_{1-x}$Cd$_{x}$Te/Cd
QW would be enlarged so one needs larger gate voltage to drive this
band insulator to the topological insulator.

\begin{figure}[t]
\includegraphics[width=1\columnwidth]{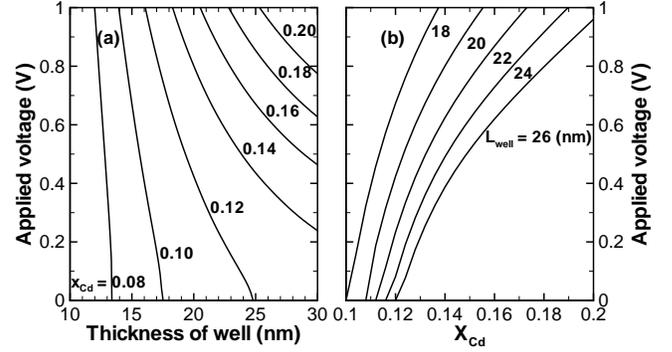}
\caption{The critical gate voltage as function of (a) the thickness
of well and (b) the composition of Cd for different
CdTe/Hg$_{1-x}$Cd$_{x}$Te/CdTe QWs at $T=0$ K.} \label{fig:fig3}
\end{figure}

In Fig. \ref{fig:fig3} (a) we plot the phase diagram as function of
the gate voltage (equivalently electric field) and the thickness of
QWs for different compositions of Cd atoms at $T=0$ K. For a given
QW with a fixed Cd composition and thickness of QW, a quantum phase
transition between a BI and a TI takes place when the applied
voltage is larger than a critical voltage. One see that higher
voltage is needed to induce the transition in narrow QWs for a fixed
composition. For a fixed thickness of the QW, the transition is more
easily induced at small compositions. Fig. \ref{fig:fig3} (b)
displays the phase diagram as function of the gate voltage and the
Cd compositions. With increasing the external voltage, the QSH
states can even be found for high composition $x$. For wider QWs,
one can see that it is more easily to drive a normal band insulator
into a topological insulator. From Fig. \ref{fig:fig3}, we find that
the thickness of well to maintain the QW in topological insulator
phase can not be less than $13$ nm when the Cd composition in
Hg$_{1-x}$Cd$_{x}$Te well is smaller than $0.08$, and hardly
possible to see the quantum phase transition driven by external
electric field since the boundary of phase diagram almost does not
change with increasing the gate voltage. Notice that a very large
electrical field can also lead to the leaky of electrons out of the
QWs because of quantum confinement of QW.

\begin{figure}[t]
\includegraphics[width=1\columnwidth]{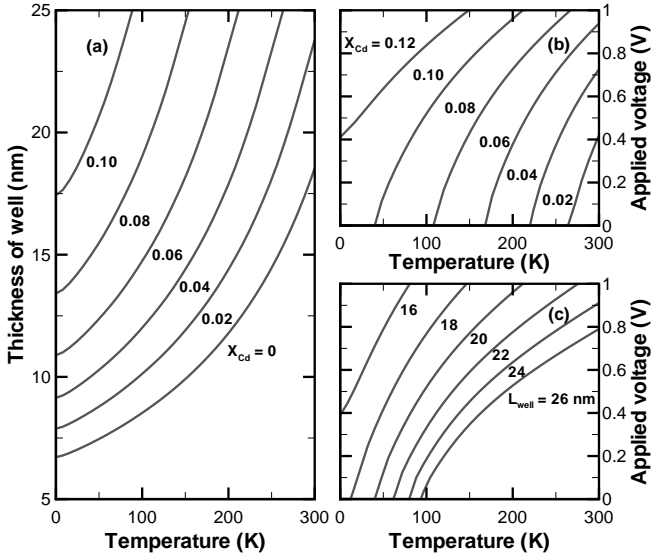}
\caption{(a) The critical thickness of well as function of
temperature for CdTe/Hg$_{1-x}$Cd$_{x}$Te/CdTe QWs with different Cd
composition without external gate voltage. (b) The critical gate
voltage as function of temperature for QWs with fixed thickness (20
nm well) and different Cd composition. The same as (b) but for QWs
with fixed Cd composition $x=0.10$ and different thickness of well.}
\label{fig:fig4}
\end{figure}

Finally, we will discuss the temperature effect on the transition
between the BI and the TI. Since the bulk bandgap of
Hg$_{1-x}$Cd$_{x}$Te varies with increasing temperature, the phase
diagram will be very different from that at $T=0$. Fig.
\ref{fig:fig4} (a) shows the critical thickness of well of the phase
transition increase rapidly (without external gate voltage) with
increasing temperature. Take the CdTe/HgTe/CdTe QW ($x=0$) for
example, the thickness of the QW varies from 6.7 nm to 18.5 nm when
temperature increase from 0 K to 300 K. Such great change indicates
that temperature is a significant factor of the phase transition
between BI and the TI in these QWs. Fig. \ref{fig:fig4} (b) and (c)
shows that the critical voltage would increase with increasing
temperature for QWs with fixed Cd compositions and thickness of
well. Decreasing the Cd composition or increasing the thickness of
well would make the phase transition possibly occurs at higher
temperature and smaller gate voltage. One can see from the panel (b)
that at T = $300$ K, it is possible to drive a
CdTe/Hg$_{1-x}$Cd$_{x}$Te/CdTe QW with $x=0.02$ and $20$ nm well
width at the voltage between the top and bottom gates $U_{ex}=0.4$
$V$.

We demonstrate theoretically that external electric field and temperature
can drive a quantum phase transition between a band insulator and a
topological insulator in CdTe/Hg$_{1-x}$Cd$_{x}$Te/CdTe quantum wells. It
provides us an efficient way to drive the band insulator into the
topological insulator. Our theoretical result is interesting both from the
basic physics and potential application of the spintronic devices based on
this novel topological insulator system.

\begin{acknowledgments}
This work was supported partly by the NSFC Grant Nos. 60525405 and 10874175,
and the bilateral program between China and Sweden.
\end{acknowledgments}


\begin{thebibliography}{99}
\bibitem{Kane} C. L. Kane and E. J. Mele, Phys. Rev. Lett. \textbf{95},
226801 (2005).

\bibitem{BHZ} B. A. Bernevig, T. L. Hughes, and S. C. Zhang, Science \textbf{%
314}, 1757 (2006).

\bibitem{Marcus} M. Konig, S. Wiedmann, C. Brune, A. Roth, H. Buhmann, L.W.
Molenkamp, X. L. Qi, and S. C. Zhang, Science \textbf{318}, 766 (2007).

\bibitem{Fu} L. Fu, C. L. Kane, and E. J. Mele, Phys. Rev. Lett. \textbf{98}%
, 106803 (2007).

\bibitem{Hsieh} D. Hsieh, Y. Xia, L. Wray, D. Qian, A. Pal, J. H. Dil, J.
Osterwalder, F. Meier, G. Bihlmayer, C. L. Kane, Y. S. Hor, R. J. Cava, M.
Z. Hasan, Science 323, \textbf{919} (2009).

\bibitem{OtherMat} H. Zhang, C. X. Liu, X. L. Qi, X. Dai, Z. Fang, and S. C.
Zhang, Nat. Phys. \textbf{5}, 438 (2009).

\bibitem{Rashba} Y. A. Bychkov and E. I. Rashba, J. Phys. C \textbf{17},
6039 (1984).

\bibitem{SpinHall} W. Yang, Kai Chang, and S. C. Zhang, Phys. Rev. Lett.
\textbf{100}, 056602 (2008).

\bibitem{Burt} M. G. Burt, J. Phys.: Condens. Matter \textbf{4}, 6651
(1992); B. A. Foreman, Phys. Rev. B \textbf{56}, R12748 (1997); T. Darnhofer
and U. Rossler, \textit{ibid}. \textbf{47}, 16020 (1993).

\bibitem{Parameters1} \emph{II-VI and I-VII Compounds; Semimagnetic Compounds%
}, edited by Landolt-B\"{o}rnstein, Group III Vol. 41B, edited by U. R\"{o}%
ssler (Springer-Verlag, Berlin, 1999).

\bibitem{Parameters2} X. C. Zhang, A. Pfeuffer-Jeschke, K. Ortner, V. Hock,
H. Buhmann, C. R. Becker, and G. Landwehr, Phys. Rev. B \textbf{63}, 245305
(2001).

\bibitem{Parameters3} C. R. Becker, V. Latussek, A. Pfeuffer-Jeschke, G.
Landwehr, and L. W. Molenkamp, Phys. Rev. B \textbf{62}, 10353 (2000).

\bibitem{SpuriousSolution} W. Yang and Kai Chang, Phys. Rev. B \textbf{72},
233309 (2005).

\bibitem{SOIRWinkler} R. Winkler, \emph{Spin-Orbit Coupling Effects in
Two-Dimensional Electron and Hole Systems}, (Springer-Verlag, Berlin, 2003),
Chap. 3, pp. 29-33.

\bibitem{NonlinearRSS} W. Yang and Kai Chang, Phys. Rev. B 74, 193314 (2006).
\end{thebibliography}
\end{document}